# Scheme for Low Energy Beam Transport with a non-neutralized section

A. Shemyakin**, L. Prost, Fermilab*, Batavia, IL 60510, USA

*Abstract*
A typical Low Energy Beam Transport (LEBT) design relies on dynamics with nearly complete beam space charge neutralization over the entire length of the LEBT. This paper argues that, for a beam with modest perveance and uniform current density distribution when generated at the source, a downstream portion of the LEBT can be un-neutralized without significant emittance growth.

## 1. Introduction

A Low Energy Beam Transport (LEBT) line, the part of a modern ion accelerator between an ion source (IS) and a Radio-Frequency Quadrupole (RFQ), may serve several purposes, which include matching the optical functions between the IS and RFQ, providing differential pumping to improve the RFQ vacuum, shaping the beam time structure with a chopping system, and assisting in the selection of the right ion species. A typical design includes 1-3 solenoidal lenses for focusing and relies on transport with nearly complete beam space charge neutralization over the entire length of the LEBT. However, neutralization is inevitably broken in some part of the LEBT if a chopper creates beam pulses. As a result, the pulse front may have Twiss parameters significantly different from their steady state values, creating beam losses in the downstream accelerator. Shortening the transition time by increasing the vacuum pressure is often in contradiction with the pursuit for high reliability of the adjacent RFQ.

In this paper, we discuss the possibility and rationality of imposing un-neutralized (i.e. space charge dominated) beam transport dynamics in a portion of the LEBT. For estimations, we will use the parameters from PXIE [1], a test accelerator presently being constructed at Fermilab, and assume that the beam ions are negative hydrogen (H$^-$).

## 2. Reasoning for a LEBT with inclusion of an un-neutralized section

Often, a LEBT either operates with a pulsed ion source or is capable of creating pulses from an initially DC beam. Because the ionization process is not instantaneous, the front of the beam pulse is not neutralized as it propagates through the LEBT. Thus for long-pulse operation, when the accelerator optics is tuned for neutralized transport in the LEBT, the space charge at the beginning of the pulse may result in increased losses in the LEBT proper and the following beam line (e.g.: RFQ). In the case of PXIE, several microseconds pulses envisioned for tuning will not provide a representative envelope for CW operation.

Remedies include working at relatively high pressure to speed up the neutralization process, which, in turn shortens the time for the beam parameters to reach a steady state, and moving the chopping system as close as possible to the RFQ in order to decrease the distance that beam travels with full space charge and low energy. In both cases, reliability of the RFQ may suffer because of higher pressure and/or increased irradiation of the RFQ vanes during chopping.

Here we consider an alternative scheme, where the ion source works in the DC mode and the beam propagates through the first, 'high pressure' part of the LEBT being neutralized, but neutralization is stopped right upstream of an electrostatic chopper. Hence, in the ideal case with zero density for the neutralizing ions in the downstream part of the LEBT, the beam envelope is time-independent.

Applicability of such scheme depends on several factors, most importantly, the beam perveance

$$P_b = \frac{I_b}{U_{IS}^{3/2}} \qquad (1)$$

where $I_b$ is the beam current and $U_{IS}$ is the ion source bias voltage. If the perveance exceeds a certain limit, an un-neutralized beam simply cannot be transported in a LEBT with lumped focusing even in the linear approximation. In addition, even for a lower perveance, non-linear space charge effects can dramatically increase the beam emittance, making it not suitable for an accelerator.

## 3. Linear space charge effects

To estimate the maximum perveance $P_{bm}$ theoretically allowing lumped focusing, let's consider the space-charge dominated transport of a non-relativistic, round, completely un-neutralized H⁻ beam with uniform transverse distribution of the charge density, i.e. neglecting the beam emittance and potential drop across the beam.

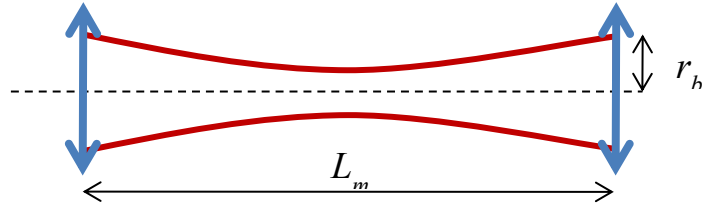

Figure 1. Model of a beam propagating between two thin lenses.

The maximum length $L_m$ that the beam can propagate between two thin focusing elements remaining within radius $r_b$ (Fig. 1) can be expressed (using, for example, formulae from Ref. [2]) as

$$\frac{L_m}{r_b} = \sqrt{2.33\frac{\sqrt{\frac{2e}{M_i}}}{P_b}} = \sqrt{\frac{3.59\,\mu A/V^{3/2}}{P_b}}. \qquad (2)$$

where $e$ is the electrical charge and $M_i$, the ion mass. In real-life, for a typical solenoid, the magnetic lens inner radius is at least twice the beam radius and its length is roughly equal to its inner diameter. One can argue that the maximum allowable perveance corresponds to the case when the minimum possible physical distance between lenses exceeds their length by only a factor of 2-3 (with a factor of '1' meaning that the lenses would be touching). Taking a factor of 3 for the lenses' distance-to-length ratio implies

$$\frac{L_{m\_sc}}{r_b} \approx 12 \tag{3}$$

and a value of the maximum allowable perveance of

$$P_{bm} \approx \left(\frac{L_{m\_sc}}{r_b}\right)^{-2} \cdot 3.59\,\mu A/V^{3/2} \approx 0.025\,\mu A/V^{3/2}. \tag{4}$$

Note that for the case of electron beams, the value in Eq. (4) merely needs to be multiplied by the square root of the H⁻-to-electron mass ratio, which leads to

$$P_{bm\_e} \approx 1\,\mu A/V^{3/2} \tag{5}$$

In the case of PXIE's LEBT, the perveance of the 10mA, 30 kV H⁻ beam is $1.9 \cdot 10^3$ μA/V$^{3/2}$, significantly lower than the estimation from Eq. (4). Therefore, for these parameters, un-neutralized transport is not excluded in this simplest model.

One can compare the latter estimation with an emittance-dominated transport. Suppose a beam with Gaussian distributions in all transverse phase space planes is transported with negligible space charge (case of a complete neutralization). One can show that the maximum distance between thin lenses where the beam rms size stays below $\sigma_{r0}$, is equal to the beta–function in the lens, or

$$L_{m\_em} = \frac{\sigma_{r0}^2}{\varepsilon} \tag{6}$$

where $\varepsilon$ is the beam rms emittance (un-normalized). For the PXIE LEBT beam with expected $\varepsilon \sim 15$ μm and typical $\sigma_{r0} \sim 7$ mm, Eqs. (2) and (6) give, correspondingly, $L_m/r_b \approx 43$ and $L_{m\_em}/(2\sigma_{r0}) \approx 230$ (factor of 2 accounts for the difference between radius of a uniform-density beam and a rms size). One can conclude that in the un-neutralized case the limitation coming from the space charge is more restrictive.

## 4. Non-linear space charge effect

A significantly more important limitation is an emittance growth due to space charge. For a beam with a Gaussian current density distribution, the space charge force is highly non-linear outside of the beam core as illustrated in Fig.2. The tail particles experience a lower radial kick, which distorts the beam phase portrait and increases the beam emittance.

An obvious solution to avoid emittance dilution due to space charge is to create a beam with uniform current density distribution in the ion source, so that the space charge force is linear. However, for any realistic beam, the thermal radial velocities affect the particles' distribution as it propagates down the beam line. It is easier to comprehend for the case where space charge is negligible and the beam initially parallel: when the thermal velocities significantly increase the beam size, as it propagates, a particle's transverse position is determined mainly by its initial radial velocity rather than its initial position. In turn, whatever the initial radial density distribution may be, it eventually becomes Gaussian, thus reflecting the thermal equilibrium in the ion source plasma.

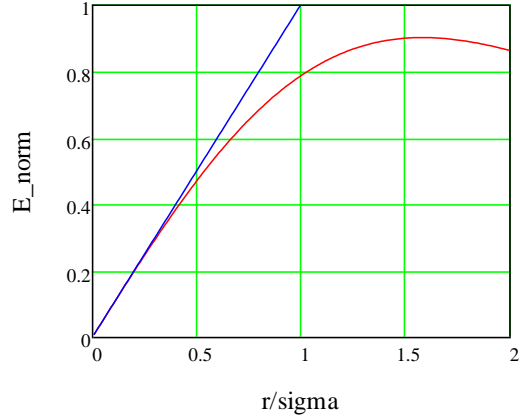

Figure 2. Normalized electric field of a beam with a Gaussian current density profile as a function of its normalized radius (red curve) compared with the field of a uniform current density beam (blue curve) with the same density at the center.

A derivation for the evolution of the current density distribution is presented in the Appendix, for the case of a beam with zero space charge, initial uniform density in space and Gaussian shape in velocities propagating through a free space. The main result is that, whether the beam shape is closer to uniform-density or to Gaussian where sampled in the beam line, is determined by a single parameter

$$\sigma_T = \frac{\sigma_{x'} L}{R_0 (1 + kL)} \qquad (7)$$

where $\sigma_{x'}$ is the initial rms velocity spread (expressed as an angle), $R_0$ is the initial beam radius, $L$ is the propagation distance, and $k \cdot R_0$ is the initial average angle at the beam boundary. The radial distribution is close to the initial uniform density at $\sigma_T \ll 0.5$ and becomes close to Gaussian at $\sigma_T \gg 0.5$ (Fig. 3).

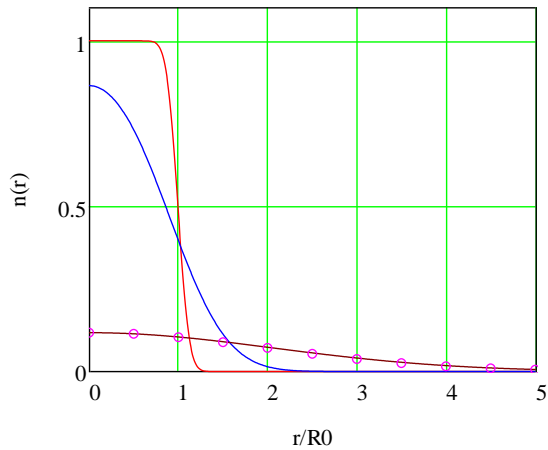

Figure 3. Normalized radial current density distribution of a beam propagating in free space with no space charge. The initial current density distribution is a step function, and the velocity profile is Gaussian (see text for details). The value of $\sigma_T$ for the solid lines is 0.1 (red), 0.5 (blue), and 2 (brown). The magenta circles show, for illustration, a Gaussian distribution.

One can consider this result from the point of view of the Courant-Snyder formalism. The phase advance for particles propagating in free space is

$$\Delta\Psi = \int_0^L \frac{ds}{\beta_0 - 2\alpha_0 s + \gamma_0 s^2} = \operatorname{atan} \frac{L}{\beta_0 - \alpha_0 L}, \qquad (8)$$

where $\alpha_0, \beta_0, \gamma_0$ are the initial Twiss parameters.

Substituting in Eq. (8) $\beta_0 = \frac{R_0^2}{4\varepsilon}$, $\alpha_0 = -k\frac{R_0^2}{4\varepsilon}$, $\varepsilon = \frac{R_0}{2}\sigma_{x'}$ leads to a simple relationship between the two descriptions

$$\tan(\Delta\Psi) = 2\sigma_T. \qquad (9)$$

Hence, $\sigma_T = 0.5$ considered previously corresponds to a phase advance of $\pi/4$.

One can use this zero-current model as guidance for designing a beam transport solution where the space charge force is initially linear. If the phase advance, calculated with space charge, is low, $\tan(\Delta\Psi) \ll 1$, most of particles should remain in the region with linear radial dependence of the space charge force. In the language of Eq.(7), the separation, $\sigma_{x'}L$, between a particle with initial rms velocity, $\sigma_{x'}$, propagating over the distance, $L$, and a particle with the same initial position but zero thermal velocity, should be much less than the radius the beam would have without thermal velocities, $R_b$, or, according to the arguments following Eq.(7),

$$\sigma_{x'} L \ll 0.5 R_b. \qquad (10)$$

Therefore, propagation with low emittance growth is limited to the length

$$L_{m\_eg} = 0.5 \frac{R_b}{\sigma_{x'}} = \frac{R_0}{4\varepsilon} R_b. \qquad (11)$$

The last equality assumes a beam with initial uniform density ($\varepsilon = \sigma_{x'} \cdot R_0 / 2$).

While the previous consideration was about propagation in free space, in most of real LEBTs the beam is focused by axially symmetrical solenoidal lenses. In the case of completely neutralized beam transport with linear optics, the lens would replicate the initial (i.e. at the ion source exit) current density distribution in the image plane, as illustrated in Figures 4 and 5. Consequently, one can re-create a beam with a uniform current density distribution but magnified with respect to the one generated at the ion source. Position of the image corresponds to the phase advance of $\pi$.

In turn, the limit distance $L_{m\_eg}$, over which the uniform-density profile is nearly preserved and, consequently, aberrations from space charge are mostly suppressed, increases in accordance with Eq.(11).

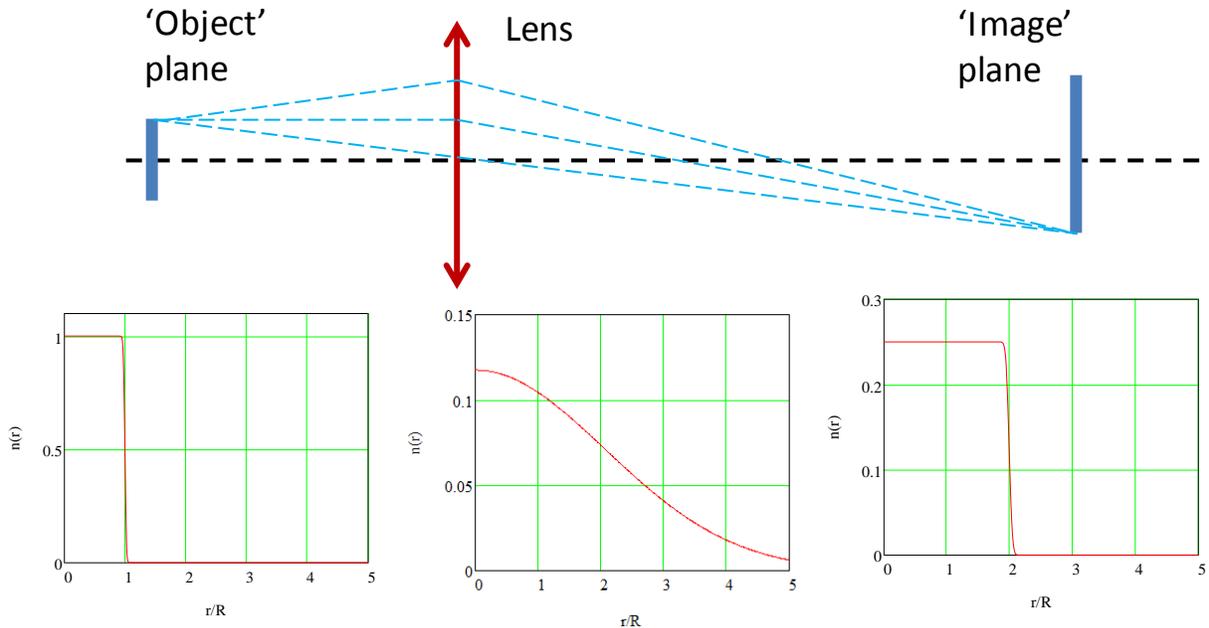

Figure 4. Illustration of the evolution of a beam with initial uniform current density distribution as it propagates through a thin focusing lens and returns to a uniform density distribution in the image plane of that lens. The plots show analytical calculations of the current density distributions in MathCad. Density and radius are normalized by the initial density in the center of the beam and the initial beam radius. The distance between the lens and the image plane is twice the distance between the emitter plane and the lens. At the lens plane $\sigma_T = 2$.

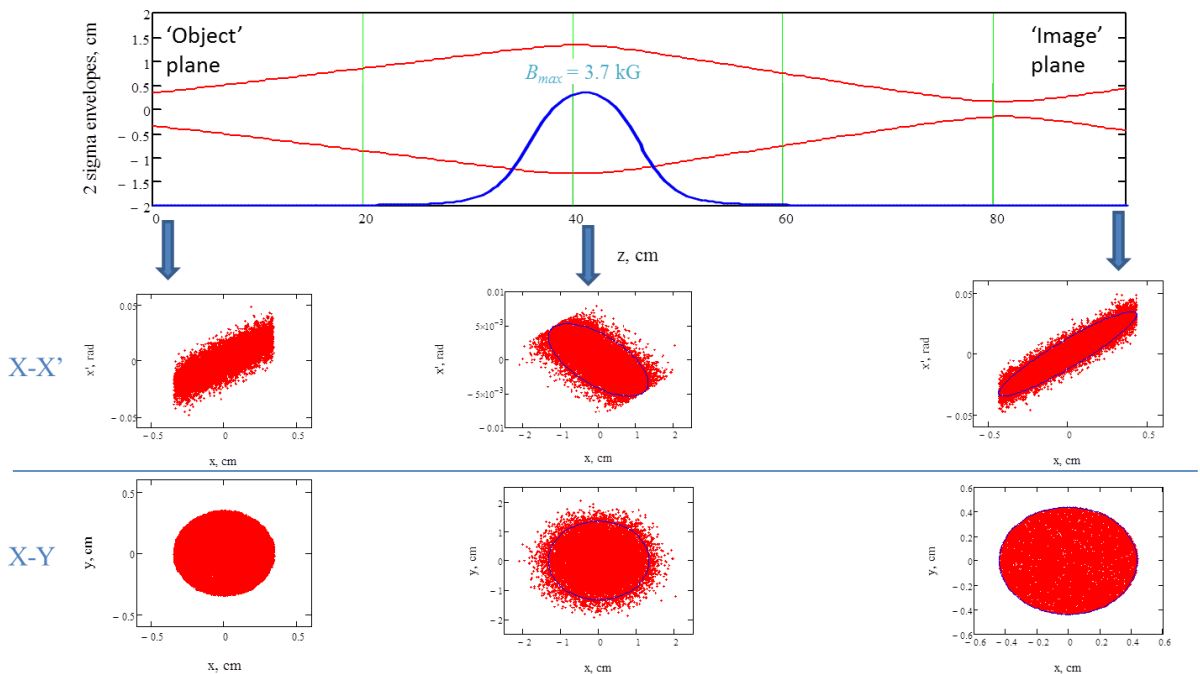

Figure 5. Numerical simulation by VACO [3] of the propagation of a beam with the initially uniform spatial density distribution and Gaussian velocity distribution through a magnetic lens.

## 5. Scheme of LEBT with partial neutralization

Based on the above considerations, the LEBT scheme with an un-neutralized section is proposed as follows:
- The ion source is optimized to generate a uniform spatial density distribution at the nominal beam current.
- Beam transport immediately following the ion source is as close as possible to being completely neutralized.
- The beam size near the image plane of the first solenoid is increased to the limit imposed by aperture limitations of critical elements downstream (e.g. chopping system).
- Near that image plane (phase advance ~ $\pi$), neutralization is interrupted: the flow of neutralizing ions from the upstream section with relatively high pressure is stopped by applying a positive voltage on an axially symmetrical electrode located near the image plane, and ions created downstream of this electrode are removed by a transverse electric field applied, for example, to the chopper's kicker electrodes.
- The phase advance over the remaining length of the LEBT is minimized, and vacuum kept as low as possible.

In this scheme, a low space charge-related emittance growth is achieved by the combination of neutralized transport in the upstream, high-pressure portion of the LEBT and the fact that space charge forces are mostly linear in the downstream part.

Such a scheme was simulated keeping in mind its practical realization at PXIE. It employs three solenoids: one to optimize the beam propagation through a space allocated for the future installation of a bend and the following two to match the beam at the RFQ entrance. A chopping system, installed between the second and third solenoids, determines the allowable beam size in this region.

To simplify simulations, we assume that the phase advance from the ion source emitting surface to the position where neutralization is complete is small and can be neglected. Consequently, the simulation begins in the location with full ion energy, complete neutralization, and uniform current density. While in a real beam line the assumption might be incorrect, the additional phase advance would result only in a shift of the optimum position of the image plane, not affecting the main conclusions.

Results of numerical simulations are presented in Fig. 6 for two cases, which differ by the initial current density distribution: uniform in one and Gaussian in another. For both cases, the velocities distribution is Gaussian. The beam $2\sigma$ envelopes are such that the Twiss parameters at the end of the beam line are nearly identical.

Fig. 6 shows that when the initial current density was chosen to be uniform, there is no emittance growth (the actual small decrease in the calculation is likely due to numerical effects), while the emittance grew by ~25% when the initial current density distribution was chosen to be Gaussian.

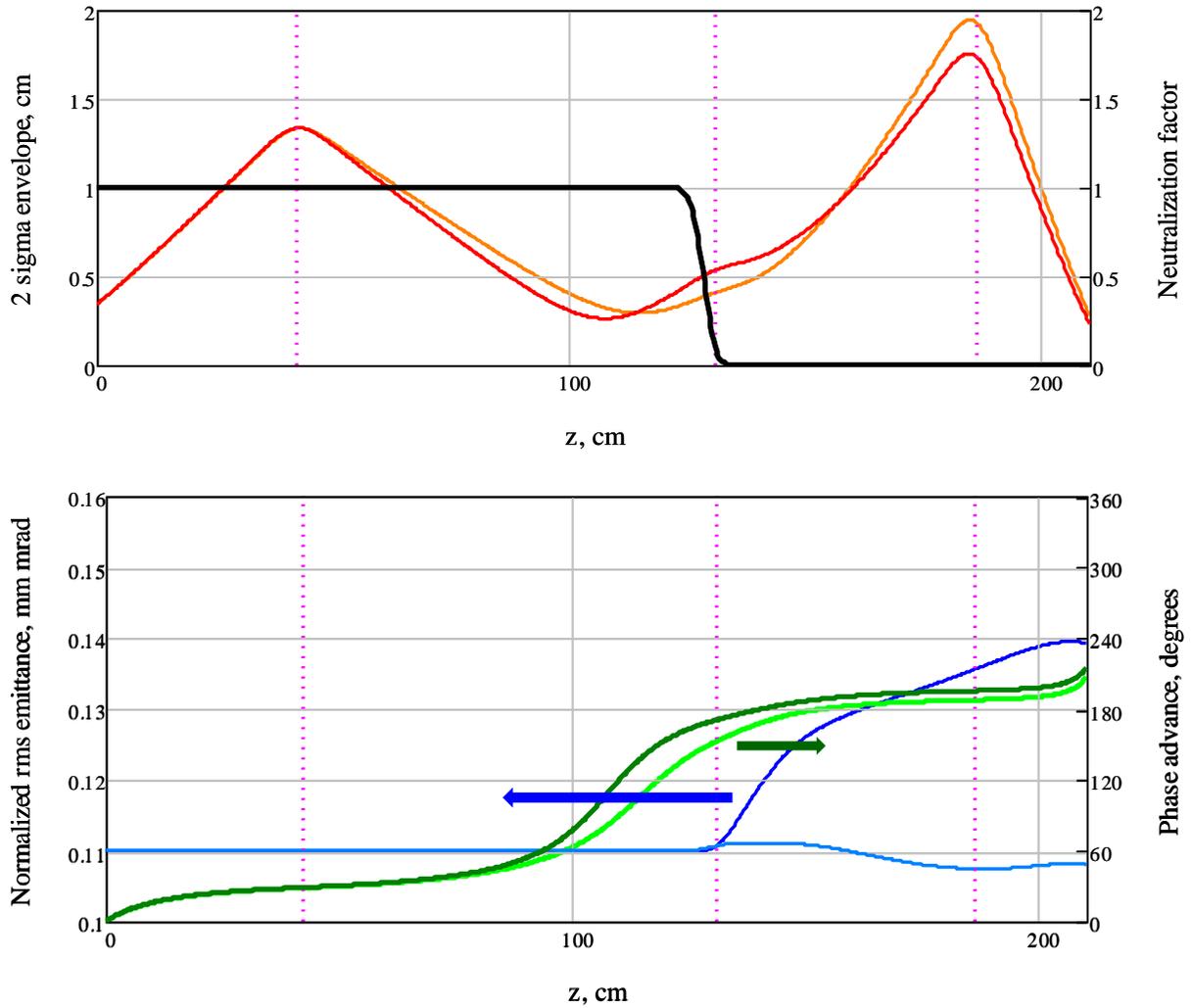

Figure 6: Top: Beam envelopes calculated with VACO (50000 macro-particles) starting with a beam distribution with uniform current density (red) and a Gaussian current density distribution (orange). The thick black line indicates the neutralization factor. Bottom: Corresponding emittance evolution (light blue for uniform current density, navy blue for Gaussian current density) and total phase advance (dark green for uniform current density, bright green for Gaussian current density). The pink dotted lines indicate the location of the solenoids (centers).

The phase advance over the non-neutralized portion of the line for the case with a uniform current density distribution input is ~0.8 rad, i.e. $\tan(\Delta\Psi) \approx 1$. Thus, according to the reasoning based on propagation in free space, the deviation from a uniform density distribution is significant and some emittance growth would be expected. Nevertheless, the emittance did not grow. Likely, Eq.(11) is too restrictive because it takes time for particles travelling in a non-linear field to accumulate non-linear perturbations and eventually lead to an increase of the emittance.

## 6. Discussion

The proposed scheme accommodates a natural pressure distribution in a LEBT line: the high pressure near the ion source, unavoidable due to the gas flow that generates the plasma in the source, helps neutralizing the upstream portion of the LEBT, while a low pressure, preferential for a reliable operation of the RFQ, is compatible with limiting the amount of neutralizing ions generated by the residual gas.

Also, in the case where an LEBT chopping system is necessary to tailor the beam pulse shape and frequency, this scheme allows locating the chopping system somewhat far from the RFQ decreasing possible detrimental effects of absorber outgassing.

Note that while, in the examples simulated above, the transition from complete neutralization of the beam to full space charge is very close to the image plane, qualitatively, the result that the emittance does no increase should not change if this transition occurs at a distance much less than $L_{m\_eg}$ from the image plane, as calculated with Eq. (11).

This concept is the base of the PXIE LEBT that is presently being commissioned at Fermilab.

## 7. Acknowledgements

Authors acknowledge stimulating discussions with J.-F. Ostiguy and J. Staples and are thankful to V. Lebedev for writing VACO.

## 9. Appendix

Let's consider the initial normalized ion velocities distribution in the form of

$$\frac{\partial^4 N}{\partial x \partial y \partial x' \partial y'} \equiv f_0(x_0, x_0', y_0, y_0') =$$

$$= \frac{1}{2\pi\sigma_{x'}^2} \frac{1}{\pi R_0^2} \exp\left(-\frac{(x_0' - kx_0)^2 + (y_0' - ky_0)^2}{2\sigma_{x'}^2}\right) \cdot \begin{cases} 1, & \sqrt{x_0^2 + y_0^2} \leq R_0 \\ 0, & \sqrt{x_0^2 + y_0^2} > R_0 \end{cases} \quad \text{(I)}$$

where $k$ characterizes initial beam divergence. After beam travels over the distance $L$, the current density $j_1$ at the location $(x_1, y_1)$ is determined by integration of all trajectories satisfying

$$x_1 = x_0 + x_0'L; \quad y_1 = y_0 + y_0'L. \quad \text{(II)}$$

One can replace integration over initial angles by integration over initial coordinates according to Eq.(II):

$$j_1(x_1, y_1) = I_b \int dx_0' \int dy_0' f_0(x_0, x_0', y_0, y_0') = \frac{I_b}{L^2} \int dx_0 \int dy_0 f_0(x_0, x_0', y_0, y_0') =$$

$$= \frac{1}{2\pi\sigma_{x'}^2} \frac{I_b}{\pi R_0^2} \frac{1}{L^2} \int_{-R_0}^{R_0} dx_0 \int_{-\sqrt{R_0^2 - x_0^2}}^{\sqrt{R_0^2 - x_0^2}} dy_0 \exp\left(-\frac{\left(\frac{x_1 - x_0}{L} - kx_0\right)^2 + \left(\frac{y_1 - y_0}{L} - ky_0\right)^2}{2\sigma_{x'}^2}\right) = \quad \text{(III)}$$

$$= \frac{1}{2\pi\sigma_{x'}^2} \frac{I_b}{\pi R_0^2} \frac{1}{L^2} \int_{-R_0}^{R_0} dx_0 \exp\left(-\frac{\left(x_0 - \frac{x_1}{1+kL}\right)^2}{2\sigma_1^2}\right) \int_{-\sqrt{R_0^2 - x_0^2}}^{\sqrt{R_0^2 - x_0^2}} dy_0 \exp\left(-\frac{\left(y_0 - \frac{y_1}{1+kL}\right)^2}{2\sigma_1^2}\right)$$

where $\sigma_1 \equiv \frac{\sigma_{x'} L}{1+kL}$ and $I_b$ is the beam current. Because of axial symmetry, $j_1$ is a function of only radius, and one can limit consideration to the case of $y_1 = 0$ ($x_1 \to r_1$):

$$j_1(r_1) = \frac{1}{\sqrt{2\pi}\sigma_1} \frac{I_b}{\pi R_0^2} \frac{1}{(1+kL)^2} \int_{-R_0}^{R_0} dx_0 \exp\left(-\frac{\left(\frac{r_1}{1+kL} - x_0\right)^2}{2\sigma_1^2}\right) erf\left(\sqrt{\frac{R_0^2 - x_0^2}{2\sigma_1^2}}\right) \quad \text{(IV)}$$

To look at the shape of the distribution, it is convenient to normalize the variables by their values in absence of the angular spread and to introduce a new integration variable

$$j_n \equiv \frac{j_1}{I_b} \pi R_0^2 (1+kL)^2; \quad r_n \equiv \frac{r_1}{R_0(1+kL)}; \quad x_0 \to t \equiv \frac{x_0}{R_0}. \quad \text{(V)}$$

After that, the shape of the distribution is defined by a single dimensionless parameter

$$\sigma_T \equiv \frac{\sigma_1}{R_0} = \frac{\sigma_{x'} L}{R_0(1+kL)} \quad \text{(VI)}$$

as follows

$$j_n(r_n) = \frac{1}{\sqrt{2\pi}\sigma_T} \int_{-1}^{1} dt \exp\left(-\frac{(t-r_n)^2}{2\sigma_T^2}\right) erf\left(\sqrt{\frac{1-t^2}{2\sigma_T^2}}\right). \tag{VII}$$

The shape of the current density distribution can be characterized by the rms beam width. The simplest way to calculate it is by using in the integration the current density distribution as formulated in Eq. (III)

$$\sigma_{x_1}^2 \equiv \frac{1}{I_b} \int_{-\infty}^{\infty} x_1^2 dx_1 \int_{-\infty}^{\infty} j_1(x_1, y_1) dy_1 = \frac{1}{2\pi\sigma_{x'}^2} \frac{1}{\pi R_0^2} \frac{1}{L^2} \cdot$$

$$\cdot \int_{-R_0}^{R_0} dx_0 \int_{-\sqrt{R_0^2-x_0^2}}^{\sqrt{R_0^2-x_0^2}} dy_0 \int_{-\infty}^{\infty} x_1^2 \exp\left(-\frac{\left(x_0 - \frac{x_1}{1+kL}\right)^2}{2\sigma_1^2}\right) dx_1 \int_{-\infty}^{\infty} dy_1 \exp\left(-\frac{\left(y_0 - \frac{y_1}{1+kL}\right)^2}{2\sigma_1^2}\right) = \tag{VIII}$$

$$= \left[\frac{R_0}{2}(1+kL)\right]^2 + (L\sigma_{x'})^2 = \left[\frac{R_0}{2}(1+kL)\right]^2 (1+4\sigma_T^2)$$

For $\sigma_T \ll 0.5$ the width of the distribution coincides with the width of the beam with no thermal velocities. For the opposite case of $\sigma_T \gg 0.5$, it approaches $L\sigma_{x'}$, the width of a beam where the particles' positions are determined only by their initial thermal velocities and the distribution is Gaussian. It feels reasonable to designate $\sigma_T = 0.5$ as the boundary between these two states. Figure 3 (in Section 4) shows the normalized current density distributions for these 3 cases.